\documentstyle[amssym,emulateapj]{article}
\def\rad{R}
\def\trans{tr}
\def\erg{\hbox{erg}}

\def\cm{\hbox{cm}}
\def\sec{\hbox{s}}

\lefthead{DULLEMOND \& TUROLLA}
\righthead{BIMODAL ACCRETION DISKS}

\begin{document}

\title{A Note on Bimodal Accretion Disks}
\author{C.P. Dullemond}
\affil{Leiden Observatory, \\ P.O. Box 9513, 2300 RA Leiden, The Netherlands\\
e--mail: dullemon@strw.leidenuniv.nl}
\and
\author{R. Turolla}
\affil{Dept. of Physics, University of Padova, \\ Via Marzolo 8, 35131 Padova,
Italy \\ e--mail: turolla@astaxp.pd.infn.it}
  
\begin{abstract}
The existence of bimodal disks is investigated.  Following a simple
argument based on energetic considerations we show that stationary,
bimodal accretion disk models in which a Shakura--Sunyaev disk (SSD)
at large radii matches an advection dominated accretion flow (ADAF) at
smaller radii are never possible using the standard slim disk
approach, unless some extra energy flux is present. The same argument,
however, predicts the possibility of a transition from an outer
Shapiro--Lightman--Eardley (SLE) disk to an ADAF, and from a SLE disk to a
SSD. Both types of solutions have been found.
%
%
We show that for any
given accretion rate a whole family of bimodal SLE--ADAF and SLE--SSD
solutions can be constructed, each model being characterized by the
location of the transition radius.
\end{abstract}

\keywords{accretion, accretion disks ---  black hole physics --- 
hydrodynamics}

\section{Introduction}
Observations of black hole X--ray binaries (BHXBs, e.g.~V404 Cyg, Nova
Muscae, A0620-00, GRO J1655-40, Cyg X-1) suggest that a bimodal
accretion disk may be present in these sources (Thorne \& Price
\cite{thorneprice:1975}; Shapiro, Lightman \& Eardley \cite{sle:1976};
Narayan, McClintock \& Yi~\cite{narclintyi:1996}; Esin, McClintock \&
Narayan \cite{esinmcclnar:1997}). The inner part is a hot, optically
thin, quasi--spherical accretion flow, producing a non--thermal
spectrum via synchrotron radiation and Comptonization, with a cut--off
around 100 keV.  The outer part is a geometrically thin, optically
thick disk (Shakura \& Sunyaev \cite{shaksuny:1973}; Frank, King \&
Raine \cite{frankkingraine:1992}) which emits a multi--color blackbody
spectrum peaked around a few keV. By making plausible assumptions on
how the transition radius $R_{\trans}$ depends on the accretion rate
$\dot M$, one obtains a one--parameter family of spectra, once the
black hole mass $M$ and the viscosity parameter $\alpha$ are
specified. The different spectral states of the BHXBs can then be
explained by varying the accretion rate $\dot M= \dot m\dot M_{Edd}$,
where $\dot M_{Edd}$ is the Eddington accretion rate.

Historically the hot, inner region was first thought to be a
cooling--dominated, optically thin accretion disk, the SLE disk
(Shapiro, Lightman \& Eardley \cite{sle:1976}, henceforth
SLE). However, such disks are subject to a violent thermal
instability, which makes them useless for constructing stationary
bimodal models. An alternative to the SLE disk could be the recently
discovered hot, advection dominated accretion flows (ADAFs: Narayan \&
Yi \cite{narayanyi:1994}, \cite{narayanyi:1995}; Abramowicz et al.
\cite{abrchen:1995}), which are stable against both thermal and
viscous perturbations. ADAF models have been quite successful in
reproducing the spectral properties of BHXBs in quiescence and of the
Galactic center source Sgr A$^{*}$ (Narayan, Yi \& Mahadevan
\cite{naryimah:1995}). A bimodal disk structure, in which an outer
Shakura--Sunyaev disk (SSD) connects smoothly to the inner ADAF, seems
very promising in explaining the whole range of spectral states of
BHXBs, from the quiescent to the high state (Esin, McClintock \&
Narayan \cite{esinmcclnar:1997}).

Advection dominated accretion flows are dynamically very stable.  The
robustness of the ADAF solutions led Narayan \& Yi
(\cite{narayanyi:1995}) to suggest that perhaps nature will always
select this option whenever it becomes available. Yet, the precise
mechanism through which the SSD material is converted into an ADAF
remains a matter of debate. Observations of Cyg X-1 in the low state
suggested that the transition from a cold to a hot disk might result
from the secular instability of the radiation--pressure dominated
inner part of the SSD (SLE; Ichimaru \cite{ichimaru:1977}). The even
more violent thermal instability of the radiation--pressure dominated
region (Pringle, Rees \& Pacholczyk \cite{pringreespac:1973}; Piran
\cite{piran:1978}) might also be held responsible for such a
transition. The drawback is that the transition radius for these
models is usually quite close to the black hole, $\rad_{tr}\simeq
45\alpha^{2/21}\dot m^{16/21}m_{*}^{2/21}\rad_g$, where $m_{*}\equiv
M/M_{\odot}$ and $\rad_g = 2GM/c^2$ is the Schwarzschild radius (SLE;
Frank, King \& Raine~\cite{frankkingraine:1992}), whereas observations
of many BHXBs (e.g.~Narayan, Barret \& McClintock
\cite{narbarclint:1997}; Hameury et al.~\cite{hamlasclnar:1997}) seem
to imply a transition radius $\rad_{tr}\sim 10^4 \rad_g$. Moreover, no
self--consistent global solution for such a viscously/thermally driven
transition has been found so far, and, more probably, the thermal
instability results in a genuinely time--dependent behavior (e.g.~a
limit cycle, see Meyer \& Meyer--Hofmeister \cite{meyer2hofm:1981},
Mineshige \& Wheeler \cite{minwheel:1989}; Honma, Matsumoto \& Kato
\cite{honmatskat:1993}; Szuszkiewicz \& Miller
\cite{szuszkmiller1:1997}, \cite{szuszkmiller2:1997}), rather than in
a stationary transition to a hot flow. More promising in this respect
are the evaporation models in which matter evaporates from the SSD
forming an advection dominated corona (Meyer \& Meyer--Hofmeister
\cite{meyermeyhof:1994}; Narayan \& Yi \cite{narayanyi:1995}).

The question remains whether we understand why it seems not to be
possible to convert an SSD directly into an ADAF at a certain radius. This is
the question we address in this paper. We use a simple argument based
on energetics to predict which kind of bimodal disk structures should be
possible. We conclude that only bimodal accretion flows which are
formed by a SLE disk outside and by an ADAF or SSD inside are
allowed. We confirm our qualitative prediction by actually
constructing global models for both bimodal SLE--ADAF and SLE--SSD
solutions.
\section{The Equations}\label{sec-equations}
The system of equations governing the disk structure used in the
present investigation are basically those of slim disks (Abramowicz
et al.~\cite{abramczerny:1988}) and our approach follow closely those
used by Narayan, Kato \& Honma (\cite{narkathon:1997}) and Chen,
Abramowicz \& Lasota (\cite{chenabrlas:1997}) in investigating the
properties of global ADAF solutions.  More specifically, we assume
that the disk is stationary, axisymmetric, that the gravitational
field of the central object is described by the pseudo--Newtonian
potential $\Phi = -GM/(\rad-\rad_g)$ (Paczy\'nski \& Wiita
\cite{pacwi:1980}) and that self--gravity is completely
negligible. Contrary to Narayan, Kato \& Honma, we include radiative
cooling in our equations, using the approximate expressions for
bremsstrahlung losses given by Narayan \& Yi (\cite{narayanyi:1995}),
which are valid at any optical depth. Although we expect optically
thin bremsstrahlung cooling and gas pressure support to be adequate
for consistently modeling the SLE--ADAF, the
Shakura--Sunyaev disk in SLE--SSD models may be optically thick and
radiation pressure dominated, so we include radiation pressure in our
calculations.

By denoting the total pressure and the mass density at the equatorial
plane with $p$ and $\rho$, the corresponding vertically--integrated
quantities are given by
\begin{equation}
P      = 2 H p\, ,\qquad \Sigma = 2 H\rho \quad\quad\quad
\end{equation}
where $H$ is the half--thickness of the disk.
The pressure $p$ is the sum of gas and radiation pressure
\begin{equation}\label{eq-totpress-def}
p = p_g + p_r\, ,
\end{equation}
\begin{equation}\label{eq-gaspress-def}
p_g = \frac{{\cal R}}{\mu} \rho T
\end{equation}
where $T$ is the average temperature of electrons and ions
\begin{equation}\label{eq-temp-def}
T = \mu\left(\frac{T_i}{\mu_i}+\frac{T_e}{\mu_e}\right)
\end{equation}
and for simplicity we take $\mu_i=\mu_e=1$, $\mu=0.5$. We allow
$T_i$ and $T_e$ to deviate, but instead of using a sophisticated model
for electron--ion coupling we just assume $T_e=\min(T_i,\, 6\times
10^{9}\ \rm K)$.  Following Abramowicz et
al.~(\cite{abrchengrlas:1996}), we adopt for the radiation pressure
the expression
\begin{equation}\label{eq-radpress-def}
p_r = \frac{Q_{-}}{4c}\left(\tau + \frac{2}{\sqrt{3}}\right)
\end{equation}
where $Q_-$ is the cooling rate and $\tau$ is the total optical depth
(see below). Although in principle also magnetic pressure should be
accounted for, this effect, like all other effects of the frozen--in
$B$--field, will be ignored in this paper.  The isothermal sound speed
$c_s$ is given by
\begin{equation}
c_s^2=\frac{P}{\Sigma} = \frac{p}{\rho}
\end{equation}
and, following the standard approach, the disk thickness and the kinematic 
viscosity are expressed as 
\begin{equation}
H = \frac{c_s}{\Omega_K}\, ,
\end{equation}
\begin{equation}
\nu = \frac{2}{3}\alpha c_s H
\end{equation}
where $\Omega_K$ is the Keplerian rotational frequency, calculated using
the pseudo--Newtonian potential. Introducing the radial velocity $v_R$
and the angular velocity $\Omega$, the continuity, radial momentum,
azimuthal momentum and energy equations take the form
\begin{equation}
\dot M = -2 \pi \Sigma \rad v_\rad\, ,
\end{equation}
\begin{equation}\label{eq-radmom-balance}
-v_\rad\frac{dv_\rad}{d\rad}+ (\Omega^2-\Omega_K^2)\rad - \frac{1}{\rho}\frac{dp}{d\rad} = 
0\, ,
\end{equation}
\begin{equation}\label{eq-angmom-conservation}
\Sigma \rad v_\rad \frac{d(\rad^2\Omega)}{d\rad} = \frac{d}{d\rad}\left(
\Sigma\nu \rad^3\frac{d\Omega}{d\rad}\right)\, ,
\end{equation}
\begin{equation}\label{eq-energy-conservation}
\Sigma v_\rad\left(\frac{de}{d\rad} + p\frac{d\rho^{-1}}{d\rad}\right)
=Q_{+}-Q_{-}\, .
\end{equation}
We adopt for the specific internal energy (including radiation) an
equation of state of the form $(\gamma-1)e=p/\rho=c_s^2$ with constant
adiabatic index.  In the energy equation (eq.[\ref{eq-energy-conservation}]), 
the viscous heating $Q_{+}$ has the usual expression
\begin{equation}\label{eq-qplus-def}
Q_{+} = \nu\Sigma\left(\rad\frac{d\Omega}{d\rad}\right)^2\, ,
\end{equation}
while the cooling rate $Q_{-}$ is calculated using a bridging formula
valid both in the optically thin and optically thick regimes (Narayan
\& Yi \cite{narayanyi:1995}),
\begin{equation}\label{eq-q-thick-naryi}
Q_{-} = 8\sigma T_e^4 \left(
\frac{3\tau}{2} + \sqrt{3} + \frac{8\sigma T_e^4}{Q_{-}^{br}}\right)^{-1}\, .
\end{equation}
Here $\tau$ is the total (scattering plus absorption) depth
\begin{equation}
\tau = \tau_{es} + \tau_{abs}
\end{equation}
and
\begin{equation}
\tau_{es} = \frac{1}{2}\kappa_{es}\Sigma\, ,
\end{equation}
\begin{equation}
\tau_{abs}=\frac{Q_{-}^{br}}{8\sigma T_e^4}\, .
\end{equation}
The bremsstrahlung cooling function is expressed 
as (Narayan \& Yi \cite{narayanyi:1995})
\begin{equation}\label{eq-qbrems}
Q_{-}^{br} = 2Hq_{-}^{br} = 8.5\times 10^{25}H\rho^2 f_{ei} 
\erg\,\cm^{-2}\sec^{-1}
\end{equation}
where 
\begin{equation}
f_{ei}(\theta_e) = 4\left(\frac{2\theta_e}{\pi^3}\right)^{\frac{1}{2}} 
(1+1.781\;\theta_e^{1.34})
\end{equation}
and $\theta_e=kT_e/m_ec^2$ is the dimensionless electron temperature. 

\section{Bimodal disks: Energetics}
\label{sec-therm-trans}
The structure of accretion disks is determined by the set of
differential equations given in section \ref{sec-equations}.  In the
case of geometrically thin disks these equations become algebraic, so
that the disk properties may be evaluated locally (see e.g.~Frank,
King \& Raine \cite{frankkingraine:1992}). Although this is not true
any more for geometrically thick disks, a good estimate of their
structure may be nevertheless obtained from local considerations (see
e.g.~Abramowicz et al.~\cite{abrchen:1995}).

\begin{figure}
\epsfxsize=7.0cm\epsfysize=6.0cm
\centerline{\epsfbox{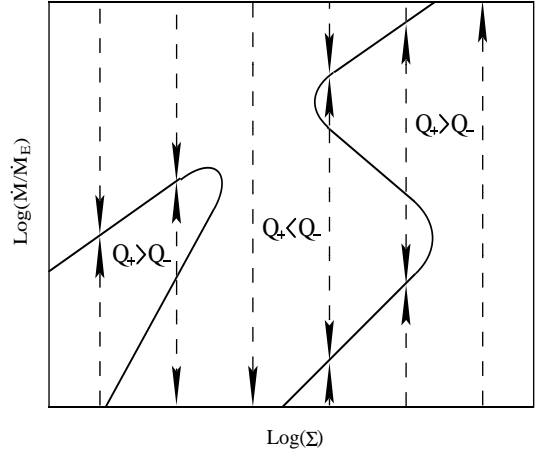}}
\caption{Unified picture of accretion flows onto black
holes. Solid lines mark the different branches of solutions. A branch
is stable when arrows point towards the solid
line. \label{fig-fig-vert}}
\end{figure}

In the analysis of the global behaviour of the models, a key role is
played by the energy equation, which simply states that viscous
heating, radiative cooling and radial energy advection must balance at
any radius
\begin{equation}\label{eq-en-loc-qq}
Q\equiv Q_{+}-Q_{-}-Q_{adv}=0\, .
\end{equation}
The non--locality of equation (\ref{eq-en-loc-qq}) is contained in
both $Q_{adv}$ and $Q_{+}$, which are functions of the gradients. For
a standard thin disk, however, $Q_{adv}\simeq 0$ and $Q_+\propto\dot
m$, while for the other classes of disks they can be reasonably well
approximated in terms of local variables. With
this proviso, the energy equation, at any given radius, may be
written as a function of $\dot m$ and $\Sigma$: $Q=Q(\Sigma, \dot
m)$. The energetically permitted configurations are then given by the
intersection of the surface $Q(\Sigma, \dot m)$ with the $Q=0$ plane,
as shown in figure \ref{fig-fig-vert} (see e.g.~Abramowicz et
al.~\cite{abrchen:1995}; Chen et al.~\cite{chenabrlas:1995}). The
curve on the left corresponds to ADAFs (upper branch) and SLE disks
(lower branch), the S--shaped curve to `slim disks' (upper branch),
radiation pressure supported SSD (middle branch) and gas pressure
supported SSD (lower branch).

Two of the five branches of accretion disks are well known to be
thermally unstable, as can be easily inferred from
figure~\ref{fig-fig-vert}.  The unstable branches are those for which
$Q_{+}>Q_{-}$ above and $Q_{+}<Q_{-}$ below the curve.  For a
solution on the radiation pressure supported SSD or SLE
branch, a sudden local heating (which amounts to a
vertical upward displacement in the $\log\Sigma - \log\dot m$ diagram)
%
%
causes the disk to enter a region where heating exceeds cooling
($Q_{+}> Q_{-}$). This results in more heating and a runaway thermal
instability sets is. A local cooling of the unstable disk similarly
produces a runaway cooling. Although the plot of figure 1 refers to
the low viscosity case, the same conclusions hold for large values of
$\alpha\sim 0.2-1$, when the topology of the solutions in the
$\log\Sigma-\log\dot m$ plane changes (a very clear example can be
found e.g. in Abramowicz \& Lasota \cite{abrlas:1995}).

In the rest of this section we show how the $\log\Sigma - \log\dot m$
diagram can be used also to identify the possible classes of bimodal,
stationary flows, i.e. two types of disks connected by a boundary
layer. Although the following analysis is somewhat qualitative, it
leads to rather interesting conclusions which are confirmed by the
numerical calculations discussed in sections \ref{sec-sle-adaf} and
\ref{sec-sle-ssd}.  Let us consider again the energy equation. Using
the exact (differential) expression for $Q_{adv}$, but retaining a
`local' approximation for $Q_+$, equation (\ref{eq-en-loc-qq})
becomes
\begin{equation}\label{eq-stat-en-difeq}
Q_{adv} = \Sigma v_\rad T \frac{ds}{d\rad} = Q_{+} - Q_{-}
\end{equation}
where $s$ is the specific entropy. In a stationary flow $v_Rds/dR$ is
just the Lagrangian rate of change of the specific entropy of a gas
element (in the case under examination, a `vertically--integrated'
thin ring).  Suppose now that the gas parcel is located at a given
radius and that it has a specified value of $s=s_0$. Assume, just for
concreteness, that $s_0$ is such that the configuration lies a bit to
the left of the stable SSD branch of figure \ref{fig-fig-vert}, so
that it is $Q_{+} < Q_{-}$. As time elapses, viscous stresses cause
the gas parcel to move inwards and, since $\dot m$ is constant, this
amounts to a horizontal shift in the $\log\Sigma - \log\dot m$
diagram. Because cooling exceeds heating, the specific entropy must
decrease in time and this implies that the configuration moves to the
right, i.e. towards the curve.  If the element is initially to the
right of the branch, heating exceeds cooling, $s$ increases in time
and the gas parcel shifts again towards the SSD branch. One can then
conclude that, even if the flow is not in the SSD phase at a certain
radius, it will try to become a SSD for smaller radii. The same
reasoning holds both for the stable and for the unstable SSD branches,
since, in both cases, one has $Q_+<Q_-$ to the left and $Q_+>Q_-$ to the
right of the curve. A major consequence is that stationary, bimodal 
accretion flows which consists of an outer SSD and a different kind of flow
at smaller radii seem not to be allowed.

When applied to the SLE branch, however, this argument yields just the
opposite result. Now it is $Q_+<Q_-$ to the right and $Q_+>Q_-$ to the
left of the curve, so, unless it lies exactly on the SLE branch, the
spiraling--in matter moves away from the curve. Depending on the
initial value of the entropy, it will end either on the ADAF or on the
SSD branch. Bimodal, stationary SLE--ADAF and 
SLE--SSD models seem therefore possible, in the sense that they are at
least energetically permitted. 

The same conclusions may be drawn looking at the integral curves of
equation (\ref{eq-stat-en-difeq}), which are shown in figure \ref{fig-svsr}.
By integrating equation (\ref{eq-stat-en-difeq}) inwards, starting
close to the SLE branch, the solutions diverge exponentially, until
either the ADAF or the SSD branch is reached. This is the stationary
transition from a SLE to either an ADAF or a SSD. A similar transition
cannot be found starting from a SSD/ADAF and integrating inwards,
because now the solutions exponentially relax back onto the SSD/ADAF
branch.

\begin{figure}
\epsfxsize=7.0cm\epsfysize=6.0cm
\centerline{\epsfbox{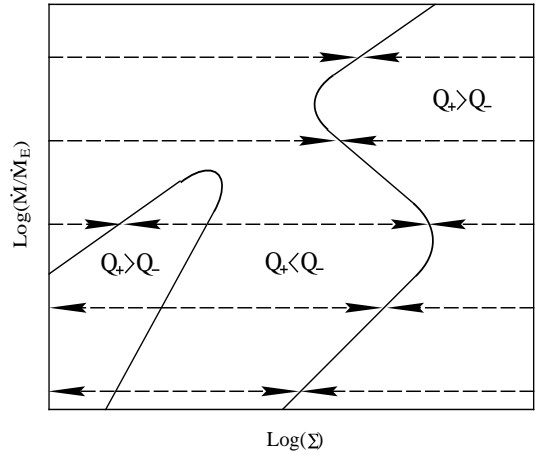}}
\caption{Same as in figure \ref{fig-fig-vert} for the 
transition analysis. Stationary, global bimodal models are possible 
for those solutions which lie on branches with outward pointing 
arrows. The only option is the SLE branch. The radiation pressure
dominated SSD branch is unstable, but cannot change 
into a hot ADAF in a stationary model.\label{fig-fig-horiz}}
\end{figure}

\begin{figure}
\epsfxsize=7cm\epsfysize=6cm
\centerline{\epsfbox{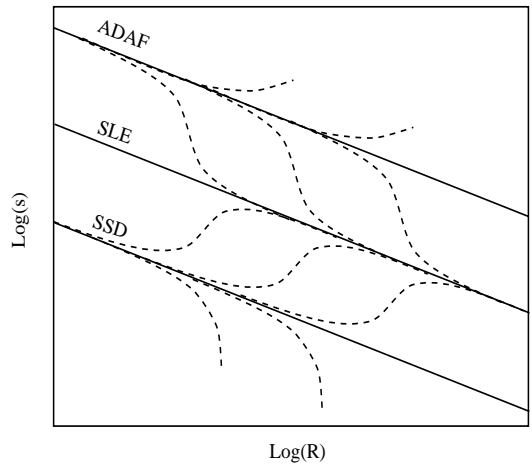}}
\caption{A schematic plot of the integral curves (dashed
lines) of equation (\ref{eq-stat-en-difeq}). The solid lines are the
ADAF, SLE and SSD branches. Each curve is found by starting at some
radius with a given specific entropy and integrating inwards. It
is clearly seen that integral curves tend to move away from the SLE
branch but converge to either the ADAF or the SSD branch.
\label{fig-svsr}}
\end{figure}

In general one can say that disks on a branch with $Q_{+}>Q_{-}$ on
its left and $Q_{+}<Q_{-}$ on its right can perform a stationary
transition to a different state towards smaller radii, while disks on
a branch with $Q_{-}>Q_{+}$ on its left and $Q_{-}<Q_{+}$ on its right
can not, as illustrated in figure \ref{fig-fig-horiz}. Much as in the
case of the stability analysis (see above), the same considerations
apply to high--$\alpha$ flows (as can be easily verified from figure 1
of Abramowicz \& Lasota \cite{abrlas:1995}), so our result is
independent of the viscosity parameter. Clearly, the inclusion of some
extra (non--local) heating/cooling process in the energy equation
changes the picture outlined above.  As shown by Honma
(\cite{honma:1996}) and Meyer \& Meyer--Hofmeister
(\cite{meyermeyhof:1994}), convection and electron conduction may
indeed provide a way to obtain global SSD--ADAF models.

The analysis presented in this section relies on the tacit assumption
that none of the branches change as we follow the evolution of a
gas element towards smaller radii. Of course this is not entirely 
true, since, at variance with what
is done in the classical stability analysis, the radial coordinate
of the gas parcel now varies because matter is spiraling in. Strictly
speaking, one should draw many curves on the $\log\Sigma - \log\dot m$ 
diagram, each referring to the different radii crossed by the gas
element.  We note, however, that the time over which the topology of
the $\log\Sigma - \log\dot m$ diagram changes is the viscous time of
radial motion. The gas element evolves on a thermal timescale which
is much shorter than the viscous time scale, so the small deformation of the
curves can be safely ignored to first order.  Much more delicate is
the assumption that $Q_+$ can be evaluated in terms of local
variables. In the transition region the derivative of $\Omega$
may deviate strongly from its usual value,
%
%
casting some doubts on
our conclusions.  A strong test for the simple argument presented
above is to actually construct the bimodal solutions for the cases in
which transitions {\em are} allowed, by solving the full set of
differential equations. Both the global SLE--ADAF and SLE--SSD
solutions have been found numerically and are presented in sections
\ref{sec-sle-adaf} and \ref{sec-sle-ssd}.
\section{Bimodal SLE--ADAF Flows}\label{sec-sle-adaf}
In this section we present our numerical models for bimodal SLE--ADAF
disks. We construct solutions which are in the ADAF phase for small
radii and in the SLE phase for large radii. We shall show that the
radius $\rad_{tr}$ at which the transition occurs is a free parameter,
with the only limitation that it must be $\rad_{tr}<\rad_{max}$. The
existence of a maximum transition radius follows from the fact that,
for too large radii, radiative cooling becomes too strong to allow of
a hot, optically thin solution. A rough estimate of $\rad_{max}$ may
found by equating the viscous heating to the optically thin 
bremsstrahlung cooling
(eqs. [\ref{eq-qplus-def}] and [\ref{eq-qbrems}]) and giving
reasonable guesses of the values of $v_{\rad}(\rad), c_s(\rad)$ and
$\Omega(\rad)$ for an ADAF (Abramowicz et al.~\cite{abrchen:1995}),
\begin{equation}
\rad_{max} \sim 2\times 10^4\frac{\alpha^4}{\dot m^2}\rad_g\, .
\end{equation} 
Sufficiently within this radius bimodal stationary solutions, which
are SLE for $\rad_{tr}<\rad<\rad_{max}$ and ADAF for $\rad<\rad_{tr}$,
exist.

Equations (\ref{eq-radmom-balance}, \ref{eq-angmom-conservation},
\ref{eq-energy-conservation}) form a fourth order set of differential
equations. We write the second order angular momentum equation
(\ref{eq-angmom-conservation}) as two first order equations and solve
the resulting four coupled differential equations numerically using a
Henyey--type relaxation scheme (Nobili \& Turolla
\cite{nobtur:1988a}).  Since we do {\em not} use the first integral of
equation (\ref{eq-angmom-conservation}), the specific angular momentum
at the inner edge, $\ell_0$, does not appear explicitly as an
eigenvalue (see e.g.~Narayan, Kato \& Honma \cite{narkathon:1997}), but
follows from the solution once suitable boundary conditions have been
specified. Logarithmic variables are adopted throughout: $\ln r$,
where $r=2R/R_g$, $\ln(v_R/c)$, $\ln(c_s/c)$, $\ln(\Omega/\Omega_K)$
and $d\ln\Omega/d\ln r$.

Global ADAF models are transsonic (see e.g.~Narayan, Kato \& Honma
\cite{narkathon:1997}; Chen, Abramowicz \& Lasota
\cite{chenabrlas:1997}) and we expect also the bimodal solutions we
are looking for to pass through a critical point close enough to inner
rim. The location of the sonic point is an eigenvalue of the problem
and follows once all boundary conditions are specified. The Henyey
solver we use is specifically designed to locate critical points
automatically, looking for the points where the Jacobian of the system
changes sign. When a critical point is met, the solver replaces one of
the boundary condition with a regularity condition.  Like in all
relaxation--type schemes, a trial solution must be supplied to start
the iteration. Usually its exact form is not very important, the only
strong requirement being that the initial guess is transsonic. In the
present case, we first generated a global ADAF model, using a modified
self--similar solution as first guess, and then used it as the trial
solution for the bimodal disk.

Although different values of the disk parameters were tested, models
discussed here were obtained for $\alpha =0.1$, $\gamma =1.5$, $\dot m
= 10^{-3}$ and $m_* = 10$.  The inner and outer edges of the disk are
located at $\rad_{in} =1.5\rad_g$ and $\rad_{end}=10^6\rad_g$,
respectively. An equally spaced logarithmic radial grid was used with
$N\sim 300$ points. Grid--refinement was used to obtain better spatial
resolution in the transition region.

\subsection{The Boundary Conditions}\label{bc-adafsle}
In order to clarify the choice of the boundary conditions, let us
start discussing how boundary conditions should be placed for a global
ADAF solution in our scheme. Since the structure equations translate
into four first order ODEs, four conditions are needed. As discussed
above, the Henyey solver automatically puts one of these conditions (a
regularity condition) at the critical point, which is automatically
located. If a fine enough grid is used, any constraint which forces
the derivatives to stay finite at the sonic radius will do.  The
remaining three conditions can be placed at either $\rad_{in}$ or
$\rad_{out}$. As discussed e.g.~by Narayan, Kato \& Honma
(\cite{narkathon:1997}) the inner boundary condition must require the
specific angular momentum, $\ell$, to behave physically. At
$\rad_{in}$, we ask that $d\ell/d\rad =0$, which is equivalent to
$d\ln\Omega/d\ln \rad =-2$. The last two conditions are given at the
outer edge, specifying the values of $c_s$ and $\Omega$ at
$\rad_{out}$ (see again Chen, Abramowicz \& Lasota
\cite{chenabrlas:1997} and Narayan, Kato \& Honma
\cite{narkathon:1997}).

For the bimodal solutions we investigate in this paper, the outer
disk is of the SLE type. Such a disk allows only one boundary
condition at the outer edge, and the remaining condition must be
specified elsewhere. This is a somewhat subtle point which will be
discussed below.
\subsection{The Solutions}
In order to compute the SLE--ADAF solutions we start from a global
ADAF model with very small $\dot m$. 
At the outer edge the two `standard' boundary conditions
\begin{eqnarray}
\Omega(\rad_{out}) &=& \omega\Omega_K(\rad_{out})\\
c_s(\rad_{out}) &=& \sigma \Omega_K(\rad_{out}) \rad_{out} 
\label{eq-outerbc-cs}
\end{eqnarray}
are prescribed, where $0\leq\sigma\leq 1$ and $0\leq\omega\leq 1$ are
two arbitrary dimensionless constants. In the following, only the case
$\omega=1$ is considered for the sake of simplicity.  Once these
global ADAF solutions have been computed, $\dot m$ is gradually
increased up to $\sim 10^{-3}$, each time using the previous model as
the input to the next run. For $\rad_{out}=10^6\rad_g$ this does
still not influence the solutions significantly, provided that $\sigma$ remains
close to 1. Starting from these solutions, models with smaller
$\sigma$ are calculated. We find, however, that, owing to the
increasing importance of bremsstrahalung cooling, no solution exists
below a certain value of $\sigma$, which depends on $\dot m$ and
$\rad_{out}$.  As soon as this happens, we move the outer boundary
condition for $c_s$ (eq. [\ref{eq-outerbc-cs}]) to a smaller radius
$\rad_{tr}<\rad_{out}$,
\begin{equation}
c_s(\rad_{\trans}) = \sigma \Omega_K(\rad_{\trans}) \rad_{\trans} 
\end{equation}
using at the same time a value of $\sigma$ just above `critical'
($\sigma\sim 0.3$ typically).  One of the two outer boundary
conditions (eq. [\ref{eq-outerbc-cs}]) is therefore transformed into
an `internal' condition (a constraint which is placed somewhere within
the computational domain), which our solver can handle very easily.
For any value of $\rad_{\trans}$ we find a global solution which turns
out to be an ADAF for $\rad\lesssim \rad_{\trans}$ and a SLE disk for
$\rad\gtrsim \rad_{\trans}$. We thus obtain a family of SLE--ADAF
solutions parametrized by the value of the transition radius
$\rad_{\trans}$.

\begin{figure}
\epsfxsize=8.2cm\epsfysize=8cm
\centerline{\epsfbox{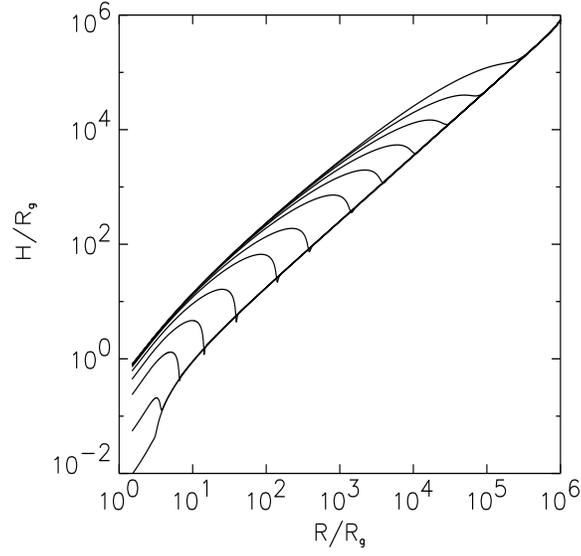}}
\caption{The disk geometry for the family of SLE--ADAF
models with $\dot m = 10^{-3}$, $m_{*}=10$, $\alpha=0.1$ and
$\gamma=1.5$.  Each model is characterized by a different location of
the transition radius.\label{fig-hhseries}}
\end{figure}

Our numerical solutions are shown in figure \ref{fig-hhseries}, where
the disk height is plotted vs. radius for different values of
$\rad_{\trans}$.  Bimodal SLE--ADAF solutions have been found
previously by Igumenshchev, Abramowicz \& Novikov
(\cite{igumabrnov:1997}), but, owing to a different choice of the
boundary conditions, numerical problems prevented them from computing
the entire family of models. The reason is that all solutions are
characterized by almost the same value of $H=c_{s}/\Omega_K$ at the
outer edge and, at least those with large enough $R_{tr}$, also at the
inner edge, as can be seen from figure \ref{fig-hhseries}.  This means
that specifying $c_s$ (or $H$ or any other variable) at the outer or
the inner edge, instead of the internal condition used here, is
formally correct but numerically unstable. Since the problem becomes
nearly singular (many solutions with almost the same boundary
condition), an impossibly accurate fine tuning would be required for
many of the solutions we have found. By setting the condition at the
transition radius this problem is solved.

\begin{figure}
\epsfxsize=8.2cm\epsfysize=4.5cm
\centerline{\epsfbox{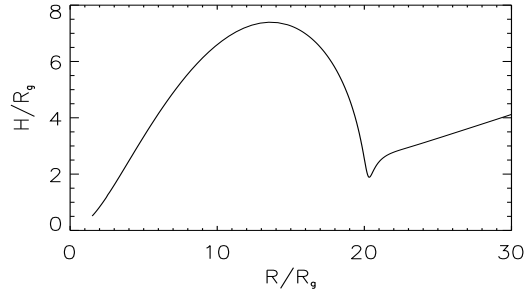}}
\caption{Close--up of the transition region for one of the SLE--ADAF
transition solutions of figure \ref{fig-hhseries}. The star marks the
position of the transition radius. \label{fig-closeup-hh}}
\end{figure}

\begin{figure}
\epsfxsize=8.2cm\epsfysize=4.5cm
\centerline{\epsfbox{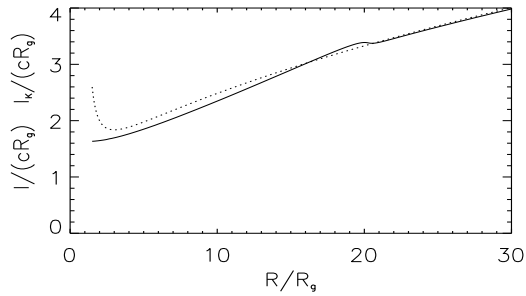}}
\caption{Same as in figure \ref{fig-closeup-hh} for
the specific angular momentum (solid line) and for the Keplerian angular
momentum (dotted line). Super--Keplerian motion is clearly visible.
\label{fig-closeup-angmom}}
\end{figure}

\begin{figure}
\epsfxsize=8.2cm\epsfysize=4.5cm
\centerline{\epsfbox{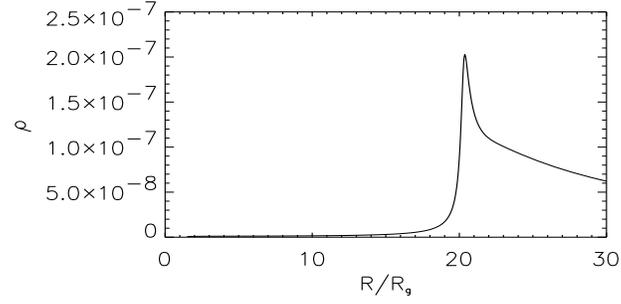}}
\caption{Same as in figure \ref{fig-closeup-hh} for
the density (in cgs units). \label{fig-closeup-dens}}
\end{figure}

\begin{figure}
\epsfxsize=8.2cm\epsfysize=4.5cm
\centerline{\epsfbox{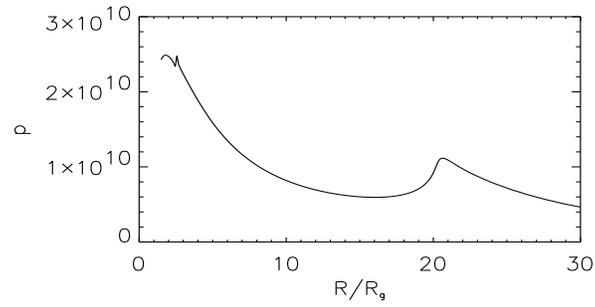}}
\caption{Same as in figure \ref{fig-closeup-hh} for the
pressure (in cgs units). The pressure drop as the SLE material enters
the ADAF is clearly seen. The small wiggle at the critical point is a
numerical artifact and is of no significance.\label{fig-closeup-press}}
\end{figure}

A close--up of the transition region is shown in figures
\ref{fig-closeup-hh}, \ref{fig-closeup-angmom}, \ref{fig-closeup-dens}
and \ref{fig-closeup-press}. Details in the behavior of the flow
variables in the transition region are of importance inasmuch they can
help in shedding light on the physical processes responsible for the
transition itself. One of the peculiar features of the transition
region is the small peak in the angular velocity which is seen in
figure \ref{fig-closeup-angmom}. 
%
%
For sufficiently small $R_{\trans}/R_{max}$,
as in the model considered here, super--Keplerian
rotation may set in. A similar behavior has been found by Honma
(\cite{honma:1996}) in modeling a SSD--ADAF transition with convective
flux, and by Igumenshchev, Abramowicz \& Novikov
(\cite{igumabrnov:1997}). Presumably the super--Keplerian rotation may
be attributed to the pressure drop experienced by the gas as it enters
the ADAF (Abramowicz, Igumenshchev \& Lasota \cite{abrigumlas:1997}).
\section{Bimodal SLE--SSD Flows}\label{sec-sle-ssd}
In this section we present numerical solutions for bimodal SLE--SSD
models. These are disks which are of SLE type for large radii and of
Shakura--Sunyaev type for smaller radii. The choice of the parameters
is the same as for the solutions discussed in the previous section,
$\alpha=0.1$, $\gamma=1.5$, $\dot m=10^{-3}$ and $m_*=10$, although
we also tried other values.  As in the SLE--ADAF
case the transition radius is arbitrary, as long as it is
$\rad_{\trans}<\rad_{max}$.  Now, however, we do not construct the
full transsonic solutions, and take the inner radius sufficiently far
away from the black hole that relativistic corrections are not
essential for the flow behavior.

The numerical approach is similar to that outlined in section 
\ref{sec-sle-adaf}, but now we use the angular momentum equation 
(\ref{eq-angmom-conservation}) in its integrated form,
\begin{equation}
\nu\rad^2\frac{d\Omega}{d\rad}=v_{\rad}(\Omega\rad^2-\ell_0)\, .
\end{equation}
This reduces the number of first order differential equations from
four to three. Since $\rad_{in}\gg\rad_g$, the specific angular
momentum at the inner edge can be assumed to vanish, $\ell_0=0$. We
verified that any sufficiently small value is allowed, and found that
the results are not very dependent on it.

The SSD part of our bimodal SLE--SSD solutions is optically thick and
may be radiation pressure supported. Since the dynamical equations
(\ref{eq-radmom-balance}, \ref{eq-angmom-conservation},
\ref{eq-energy-conservation}) contains the total pressure, rather than
the gas pressure, it is not straightforward to compute the gas
temperature $T$ from a given value of $p$ (or $c_s$).  In order to get
$T$, equations (\ref{eq-totpress-def}, \ref{eq-gaspress-def},
\ref{eq-temp-def}, \ref{eq-radpress-def}, \ref{eq-q-thick-naryi}) were
solved in a separate iteration loop at each radius for a given value
of the total pressure.
\subsection{The Boundary Conditions}
In contrast to what was discussed in section \ref{bc-adafsle}, the
number of boundary conditions now is three and, since the solution is
not transonic, none of them needs to be specified at a critical point
this time.  At the inner edge we want the flow to be a SSD, so we ask
that the radial velocity there attains the value predicted by the
standard disk model
\begin{equation}
v_\rad(\rad_{in}) = -
\frac{\alpha c_s^2(\rad_{in})}{\Omega_K(\rad_{in}) \rad_{in}}\, .
\end{equation}
As in the SLE--ADAF case, one boundary condition is given at the
outer edge, where the flow is in the SLE phase, and, again, we take 
\begin{equation}
\Omega(\rad_{out})=\Omega_K(\rad_{out})\, .
\end{equation}
The remaining condition, the one on $c_s$, needs
to be specified at the radius where we choose the transition to occur.
\subsection{The Solutions}
\begin{figure}
\epsfxsize=8.2cm\epsfysize=4.5cm
\centerline{\epsfbox{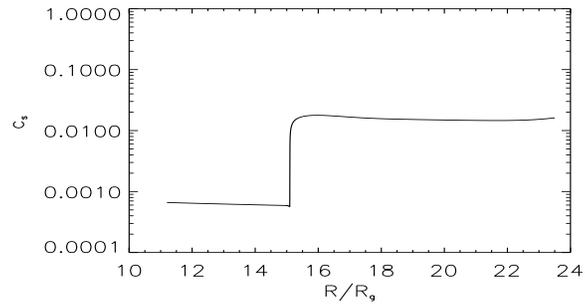}}
\caption{The isothermal sound speed (in units of $c$) for a bimodal
SLE--SSD configuration. The star marks the position of the internal
boundary condition $\rad_{\trans}$.  The solution is in the SLE phase
for $\rad>\rad_{\trans}$ and in the SSD phase for
$\rad<\rad_{\trans}$.  \label{fig-sle-ssd-cs}} 
\end{figure}

The way we actually obtained the bimodal SLE--SSD solution is similar
to that described in section \ref{sec-sle-adaf}. We started out with a
global SLE solution and then worked towards the bimodal solution by
slowly modifying the inner boundary condition on the sound speed. By
doing so we ran into the same problems as in the SLE--ADAF case, which
could only be solved by moving the condition on $c_s$ to a point
inside the computational domain, transforming it into an internal
condition.  The radius at which this condition is specified is again
the transition radius.  In the present case, however, the numerical
integration is much more troublesome because the equations become
extremely stiff in the very thin transition region and in the
SSD. This is not surprising since the cooling length in the SSD is
extremely small, and in the transition region it is even smaller.  To
circumvent this problem we slightly modified the energy equation to
guarantee 
an artificial lower limit to the cooling length. We believe that this
does not invalidate our results, as long as the cooling length remains
the smallest length scale of the problem.  A particular solution is
shown in figure \ref{fig-sle-ssd-cs}.  Contrary to the SLE--ADAF case,
now it is very difficult to obtain solutions with a different value of
$\rad_{\trans}$, although possible. For this reason, and also because
of the little practical significance of these solutions, we just
present here one model.
\section{Discussion and Conclusions}
The goal of this investigation has been to shed some light on the
physics of disk transitions. Using a simple argument based on
energetics, we have shown that global, stationary bimodal solutions
are possible if the outer disk satisfies a simple condition. Our main
result is that only an SLE disk can match an inner ADAF (or SSD),
while a bimodal accretion configuration formed by an outer SSD and an
inner ADAF is not permitted, at least within the standard slim disk
physics.

The two allowed bimodal structures, an outer SLE disk glued to an
inner ADAF or SSD, have been computed numerically. We found that, in
both cases, the transition radius turns out to be arbitrary, so a
whole family of solutions exists. This may sound odd, since there is
no physical reason to ask the sound speed to take a precise value at
an intermediate radius, as we did in finding our numerical
solutions. In principle the condition should be placed at the outer
edge. In doing so, however, one should be extremely accurate in fixing
$c_s(\rad_{out})$ to find the transition at the desired radius since
most bimodal solutions correspond to very nearly identical values
of $c_s$ (or any other variable) at the outer or inner edge.  It
remains nevertheless true that the structure of bimodal disks is
determined once a complete set of conditions is prescribed at the
edges and at the sonic radius, without the need of any extra piece of
physics.

The existence of the SLE--ADAF and SLE--SSD bimodal solutions
agrees with the energy argument of section \ref{sec-therm-trans}, and
strengthens its plausibility.  The same argument indicates that a
SSD--ADAF model is not allowed. There is a caveat here,
however. The essence of the energy argument lies in recognizing that
the non--locality of the advective cooling term in the energy equation
can be removed by performing a transformation to the fluid frame, and
doing the thermal analysis in the frame comoving with the fluid. If,
on the other hand, thermal conduction is introduced, the energy
equation becomes a second order diffusion type equation and there is
no way of removing the non--locality by changing the observer's
frame. In this case our argument definitely does not apply.  Honma
(\cite{honma:1996}) has shown that SSD--ADAF structures are possible
when the thermal heat flux from the hot ADAF into the SSD is accounted
for. In such accretion flows the location of the transition follows
self--consistently from the model, contrary to what we have found in
this paper.  This is because the heat flux will evaporate per unit
time an amount of matter from the SSD which needs to be in balance
with the actual accretion rate.  The transition radius will therefore
automatically adjust itself until this balance is reached.  In
concluding, we would like to state clearly that, because of the
thermal instability of the SLE disk, the models presented here most
probably do not exist in nature.
\begin{acknowledgements}
We wish to thank I. Igumenshchev, M.A. Abramowicz and R.
Narayan for useful comments and suggestions.
\end{acknowledgements}

\clearpage

\end{document}